\documentclass[pra,superscriptaddress,twocolumn,amsmath,amssymb]{revtex4}
\usepackage{etex}
\usepackage{amsfonts}
\usepackage{amssymb}
\usepackage{mathtools}
\usepackage{graphicx}
\usepackage{mathrsfs}
\usepackage{subfigure}
\usepackage{color}
\usepackage[normalem]{ulem}
\usepackage{natbib}
\usepackage{bbold}
\usepackage{placeins}
\usepackage{braket}
\usepackage{soul,xcolor}
\usepackage{epstopdf}
\usepackage{tabu}
\usepackage{array}
\usepackage{upgreek}
\usepackage{multirow}
\usepackage[utf8]{inputenc}
\usepackage[colorlinks = truelinkcolor = blue, urlcolor  = blue, citecolor = blue, anchorcolor = blue]{hyperref}
\hypersetup{
	colorlinks   = true, 
	urlcolor     = blue, 
	linkcolor    = magenta, 
	citecolor   = blue 
}

\begin{document}
	\title{ $\mathcal{PT}$ symmetric evolution, coherence and violation of Leggett-Garg inequalities}

\author{Javid Naikoo}
\email{javidnaikoo@gmail.com}
\affiliation{Indian Institute of Technology Jodhpur, Jodhpur 342011, India}
\affiliation{Centre for Quantum Optical Technologies, Centre of New Technologies, University of
	Warsaw, Banacha 2c, 02-097 Warsaw, Poland}
	
	\author{Swati Kumari }
	\email{swatipandey084@gmail.com }
	\affiliation{National Institute of Technology Patna, Ashok Rajpath, Patna, Bihar 800005, India}
	\affiliation{Department of Physics and Center for Quantum Frontiers of Research \& Technology (QFort), National Cheng Kung University, Tainan 701, Taiwan}
	
	\author{Subhashish Banerjee}
	\email{subhashish@iitj.ac.in}
	\affiliation{Indian Institute of Technology Jodhpur, Jodhpur 342011, India}
	
	\author{A. K. Pan }
	\email{akp@nitp.ac.in}
	\affiliation{National Institute of Technology Patna, Ashok Rajpath, Patna, Bihar 800005, India}

\begin{abstract}
We report an unusual buildup of the quantum coherence in a qubit subjected to non-Hermitian evolution generated by a Parity-Time ($\mathcal{PT}$) symmetric Hamiltonian,    which is reinterpreted as a Hermitian system in a higher dimensional space using Naimark dilation. The coherence is found to be maximum about the exceptional points (EPs), i.e., the points of coalescence of the eigenvalues as well as the eigenvectors. The nontrivial physics about EPs has been observed in various systems, particularly in photonic systems. As a consequence of enhancement in coherence, the various formulations of Leggett-Garg inequality tests show maximal  violation about the EPs.
\end{abstract}
\maketitle

\section{Introduction}
	Symmetries have played an important role in understanding and describing the physical world \cite{livio2012physics}. Their consequences include the conservation laws of physics, the existence of  degeneracies, controlling the structure of matter and dictating the interactions among fundamental particles.  An underlying symmetry in a physical system demands that the laws of physics are invariant under a particular operation. For example, the laws of physics are the same for a particle and anti-particle under the  charge conjugation ($\mathcal{C}$) operation; also for a system and its mirror image under the parity  ($\mathcal{P}$) operation and even when  time is running backward, i.e., under time reversal ($\mathcal{T}$) operation. Although symmetries are considered to be  of fundamental importance in probing the physical world \cite{weinberg2011symmetry}, it is symmetry breaking which often leads to nontrivial physics by lifting the degeneracies.

	Textbook quantum mechanics deals with Hermitian Hamiltonians having real spectra. A class of non-Hermitian Hamiltonians having real spectra endowed with an unbroken $\mathcal{PT}$ symmetry which is invariant under the simultaneous action of the parity-inversion and time-reversal symmetry operations or equivalently $[\mathcal{PT}, H] = 0$, was introduced in \cite{bender1998real}. Here, the $\mathcal{P}$ operator is defined by its action on the position ($x$) and momentum ($p$) such that under this operation $x \rightarrow -x$ and $p \rightarrow -p$. Also, $\mathcal{T}$ is an anti-linear operator, i.e., $\mathcal{T} (a_1 \Psi_1 + a_2 \Psi_2) = a_1^* \mathcal{T} \Psi_1 + a_2^* \mathcal{T} \Psi_2 $, where $\ast$ is the complex conjugation operation.  	
	The $\mathcal{PT}$ symmetric systems show a typical feature of the naturally occurring symmetries, that is, they can undergo a spontaneous symmetry breaking accompanied with real to complex transition of the eigenvalues \cite{PhysRevA.96.043810}. The points of degeneracy in $\mathcal{PT}$ symmetric systems are however very different from the conventional symmetries, in the sense that these points correspond to the coalescence of both eigenvalues as well as eigenvectors. These points are called \textit{exceptional points} (EPs). In contrast, for Hermitian Hamiltonians, the points of degeneracy are called as \textit{diabolic points} \cite{teller1937crossing} and do not involve the coalescence of eigenvectors. For a detailed account of non-Hermitian Hamiltonians in general and $\mathcal{PT}$ symmetric Hamiltonians in particular, the reader may refer to the recent reviews \cite{zyablovsky2014pt,konotop2016nonlinear,feng2017non,longhi2018parity,el2018non}.
	
	$\mathcal{PT}$ symmetric systems have been a subject matter of various studies \cite{PhysRevD.70.025001,rotter2009non,PhysRevB.63.165108,Ganainy2007,Agarwal2012spont,Agarwal2017hidden,javid2019qze,javid2019interplay,Makris,Liang2010,LJ}. In particular, the unconventional behavior of  $\mathcal{PT}$ symmetric systems around EPs has attracted considerable attention. The unidirectional reflectionless  propagation of light in photonic devices at EPs has been reported in various studies \cite{Lin2011,huang2017unidirectional}. Pronounced line broadening  around EPs in phonon laser was reported in \cite{zhang2018phonon}.  Enhanced laser performance was reported in $\mathcal{PT}$ symmetric resonators around EPs \cite{hodaei2015parity}. A striking example of nontrivial physics occurring around EPs is the reduction in light emission despite increase in pump power \cite{brandstetter2014reversing}. The enhancement of optomechanical interactions and associated nonlinearities around an EP has opened the scope for various studies \cite{PTPhononLaser}. In the context of open system dynamics, the notion of Hamiltonian EPs has been recently generalized to the Lindbladian EPs by taking into account the quantum jump operators \cite{minganti2019quantum}.
	
 Here  we bring out an interesting interplay between the $\mathcal{PT}$ symmetry (or its breaking) and the degree of coherence in a two level system. The degree of coherence for a two level system lies between zero (for maximally mixed state) and one (for pure states). Note here that for a unitary dynamics governed by the Hermitian Hamiltonian the degree of coherence of a density matrix remains unchanged due to the evolution. Coherence is at the heart of quantum interference phenomenon which plays a central role in applications of quantum theory to carry out tasks  otherwise impossible within the realm of  classical physics. Coherence is directly or indirectly responsible for all the intriguing features of quantum mechanics, viz. entanglement and  multiparticle interference which play a central role in carrying out the quantum information tasks, like teleportation \cite{bennett1993teleporting} and quantum key distribution \cite{bennett1992experimental,grosshans2003quantum}. The notion of coherence was  operationally formulated as a resource theory in \cite{Plenio201resource} and has been a theme of study of various works \cite{almeida2013probing,lloyd2011quantum,bhattacharya2016evolution,PhysRevA.91.052115,yadin2016general,dixit2019study,alok2016quantum}.  

We first demonstrate  how the degree of coherence of a qubit density matrix can be increased through the non-unitary evolution,  reinterpreted as evolution generated by a Hermitian system in a higher dimensional space using Naimark dilation,  generated by the  $\mathcal{PT}$ symmetric Hamiltonian. Here we consider the $l_1$ norm measure of the coherence which is calculated in terms of the off-diagonal elements of the density matrix. In fact, we show that the maximally mixed state $\mathcal{I}/2$ can evolve to a pure state at the EPs. As a potential application of this effect,  we examine the quantum violation of the standard Leggett-Garg  inequalities (LGIs) for testing macrorealism and find that the quantum value reaches its algebraic maximum at the EPs. Probabilistic formulations of LGIs, for example,  Wigner form  \cite{saha,pan17,swati17}, in which no explicit use of the  eigenvalues of the measured observable is required, are also studied. Since  Wigner form of LGIs are stronger than standard LGIs, their quantum violations also achieve their respective algebraic maximum. 

		The plan of this work is as follows: In Sec. \ref{sec:ModelLGIPT}, we sketch the details of the model used. Section \ref{sec:coh} is devoted to an analysis of the coherence and mixedness about the EPs. The consequences  of this effect in terms of algebraic maximum violations obtained in Leggett-Garg Inequalities is then investigated in Sec. \ref{sec:LGIandPT}. We make our conclusions in  Sec. \ref{sec:SC}.\bigskip

	\section{Model: $\mathcal{PT}$-symmetric dynamics and its extension to conventional quantum mechanics}\label{sec:ModelLGIPT}

  \subsection{$\mathcal{PT}$ symmetric time evolution}
\begin{figure}[ht]
    	\centering
    	\includegraphics[width=70mm]{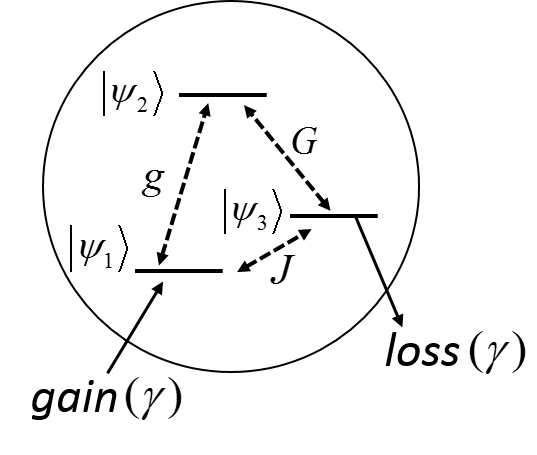}
    	\caption{(Color online) Schematic of a three level $\Lambda$ type atom. The parameters $g$, $G$ and $J$ correspond to the coupling strengths between the  levels as shown. Levels $|\psi_1\rangle$ and $|\psi_3\rangle$ are assumed to have equal gain and loss rate $\gamma$. Level $|\psi_1\rangle$ and $|\psi_3\rangle$ are connected by  radio-frequency field and simultaneously to $|\psi_2\rangle$ by two optical field modes. }
    	\label{fig:ModelPTLGI}
    \end{figure}

	In order to make the paper self-contained, we sketch a possible mechanism to generate the $\mathcal{PT}$ symmetric dynamics. A two level atom in which the two levels are coupled by a near resonant field, describes a coupled two mode system. Such a configuration can be realized experimentally in many ways. For example, two hyperfine levels can be produced by adiabatically eliminating a third intermediate level, using Raman lasers, in a $\Lambda$ type atom \cite{shahriar1990direct,Li2009em}. This configuration provides a natural example of a system with loss and gain, associated here with levels $3$ and $1$, respectively. As discussed ahead, a balance in loss and gain leads to a scenario in which  the system dynamics is governed by an effective Hamiltonian invariant under the combined operation of Parity and Time reversal.  Specifically, consider a three level $\Lambda$ type atom with the hyperfine levels represented by   $|\psi_1\rangle$,   $|\psi_2\rangle$ and  $|\psi_3\rangle$, such that $|\psi_1\rangle$ and $|\psi_3\rangle$ are coupled by an radio-frequency field and at the same time are connected to $|\psi_2\rangle$ by two optical fields, see  Fig. (\ref{fig:ModelPTLGI}). The dynamics of this system can be reduced to an effective two level system by adiabatically eliminating excited state $|\psi_2\rangle$, under large detuning condition \cite{du2018dynamical,javid2019interplay}. By assuming equal gain and loss rates, the resulting Hamiltonian is given by      
     \begin{equation}\label{eq:HPT}
     H = \begin{pmatrix}
     i \gamma           &    J \\
     J   &   - i \gamma 
     \end{pmatrix}.
     \end{equation}
     Here, $i = \sqrt{-1}$, and $\gamma$ is gain/loss parameter and $J=|1-\exp(-i\phi)|$ is the coupling strength between the two levels.
     Note that $H$ is $P$-pseudo-Hermitian, with $P =\sigma_x$, the Pauli-x operator, such that $H^\dagger = P H P^\dagger$. 
      The eigenvalues $(E_{\pm}=\pm\sqrt{J^2-\gamma^2})$ of the Hamiltonian are real for $J>\gamma$ and system is said to be in  $\mathcal{PT}$ symmetric  phase, while the case $J < \gamma$  corresponds to the $\mathcal{PT}$ symmetry broken phase. The scenario for which $J = \gamma$ represents a special case when the eigenvalues become   equal (here zero)  and the eigenvectors coalesce and is called as the \textit{exceptional point}. 
     
    \subsection{$\mathcal{PT}$ symmetry and conventional quantum mechanics}
    A system exhibiting  $\mathcal{PT}$ symmetry can be thought of as part of a larger conventional quantum mechanical system living in a higher dimensional space  by using the Naimark dilation theorem \cite{peres2006quantum}. Corresponding to the $\mathcal{PT}$ symmetric state $\ket{\psi_{PT}}$, one can construct a $\mathcal{PT}$ symmetric counterpart $\xi^{1/2} \ket{\psi_{PT}}$, and couple the system to a two level ancilla leading to the total state 
   $\ket{\psi_{total}} = \ket{\uparrow} \otimes  \ket{\psi_{PT}} + \ket{\downarrow} \otimes \xi^{1/2} \ket{\psi_{PT}}$, such that $\langle \psi_{total}| \psi_{total} \rangle =  c \langle \psi_{PT}|\eta| \psi_{PT} \rangle$ is constant under the dynamics generated by $H$. Here, $\xi = c \eta - \mathbb{1}_{2\times 2}$, $c= \sum_{i=1,2} 1/\lambda_i$, and $\lambda_i$'s are the eigenvalues of $\eta$ \cite{ueda}. The   necessary and sufficient condition for the spectrum of $H$ to be real, a scenario often called as the exact PT symmetry, is the existence of an invertible linear operator $O$, such that  $\eta = OO^\dagger$ and $H = \eta H^\dagger \eta^{-1}$ \cite{AliMost2002}. For the system under consideration, we have  
   \begin{align}
   O &= \frac{1}{J} \begin{pmatrix}
                                	i\gamma - \sqrt{J^2 - \gamma^2}           &~~~~            	i\gamma + \sqrt{J^2 - \gamma^2}\\\\
   	                                              J                                                      &~~~                     J
                                \end{pmatrix}, \nonumber \\
	\eta &= \frac{2}{J}  \begin{pmatrix}
   		                                J                  &    i \gamma\\
   		                                -i \gamma  &         J
   	                                   \end{pmatrix}.
   \end{align}
Also, the $\xi$ operator defined above turns out to be square of $\eta$. However, this special case holds only for two level systems as considered in this work.

    The unitary $U_{total} = \exp(-i H_{total} t)$, with $H_{total} = \mathbb{1} \otimes H_S + H_I$, can be constructed which governs the dynamics in the extended space. Here, $H_S$ acts locally in the system Hilbert space and $H_I$ controls the interaction between system and ancilla.  A similar scheme has been implemented   experimentally to probe the no-signaling principle under  $\mathcal{PT}$ symmetric dynamics \cite{Tang2016}.

      We now consider the time evolution of our effective two level $\mathcal{PT}$ symmetric system. The time evolution of the states $\rho_k(t) = | \psi_k (t) \rangle \langle \psi_k (t)|$ ($k=1,3$), from time $s$ to $t$ (with $t>s$), is given by the Schrodinger equation $\rho_k (t) =  U(t-s) \rho_k(s) U^\dagger(t-s)$.  It is important to note here that $U(t - s) = \exp[- i H (t-s)]$ is not a \textit{unitary} operator, since $H$ is not Hermitian. Consequently, $\rho_k (t) $ as such, is not normalized.   We define the normalized state by dividing  with the time dependent norm, i.e., 
     \begin{equation}\label{eq:rhotilde}
     \tilde{\rho}_k(t) = \frac{U(t-s) \rho_k(s) U^\dagger(t-s)}{\operatorname{Tr}\big[U(t-s) \rho_k(s) U^\dagger(t-s) \big]}.
     \end{equation}
		With $\mathcal{PT}$ symmetric Hamiltonian given in Eq. (\ref{eq:HPT}), the  time evolution operator turns out to be       
          \begin{align}\label{eq:U}
         & U(t) =   \nonumber  \\& \begin{psmallmatrix}  \tiny
                           C_h(Jt\sqrt{\alpha^2 - 1}) + \frac{\alpha S_h(Jt \sqrt{\alpha^2 - 1})}{\sqrt{\alpha^2 - 1}}      &    - \frac{i \alpha S_h(Jt\sqrt{\alpha^2 - 1}) }{\sqrt{\alpha^2 - 1}} \\
                          - \frac{i \alpha S_h(Jt\sqrt{\alpha^2 - 1}) }{\sqrt{\alpha^2 - 1}}    &    C_h(Jt\sqrt{\alpha^2 - 1}) - \frac{\alpha S_h(Jt\sqrt{\alpha^2 - 1}) }{\sqrt{\alpha^2 - 1}}    \normalsize
                                        \end{psmallmatrix}.
          \end{align}        
     Here, $C_h $ and  $S_h$  stand for the hyperbolic functions $\cosh$ and $\sinh$, respectively. Also,  $\alpha = \gamma/J$ is a dimensionless parameter, such that $\alpha = 1$ corresponds to a degeneracy in the eigenvalues of (\ref{eq:HPT}) and flags the EP. In the presence of gain/loss, i.e.,  $\gamma \neq 0$, $U(t)$ generates  a non-unitary dynamics. However, when $\gamma = 0$, the Hamiltonian becomes Hermitian leading to  unitary dynamics. In the later case, we denote  the unitary operator as $\mathcal{U} = U (t)|_{\gamma = 0}$, such that its action on  $|+\rangle \in \{|\pm\rangle = (|0\rangle \pm |1\rangle)/ \sqrt{2}\}$, is  $\mathcal{U} |+\rangle= \exp(- i Jt) |+\rangle$;  the operation $\mathcal{U}$ is incoherent in the sense  of Eq. (\ref{eq:IO}) below. However, in the non-Hermitian scenario, i.e., $\gamma \ne 0$, the (normalized) state is given by 
          \begin{equation}
          U(t) |+\rangle = \frac{1}{\sqrt{|c_1|^2 + |c_2|^2}} \big( c_1 |+\rangle + c_2 |-\rangle \big)
          \end{equation}
           with  $c_1 = \cosh( Jt \sqrt{\alpha^2 - 1}) - i \frac{1}{\sqrt{\alpha^2 - 1}}\sinh( Jt\sqrt{\alpha^2 - 1})$ and $c_2 = \frac{\alpha}{\sqrt{\alpha^2 - 1}} \sinh(Jt \sqrt{\alpha^2 - 1})$. In the limit $\alpha \rightarrow 0$, we recover the unitary dynamics and the resulting operation is incoherent.   It is worth mentioning here that, in general, the effective Hamiltonian formalism could be thought of as a semi-classical approximation. The general problem could be envisaged to be handled in terms of the quantum master equation taking into account the \textit{quantum jumps}. In such a case one would deal with the so called Liouvillian exceptional points   \cite{minganti2019quantum}.\bigskip

 \section{ $\mathcal{PT}$ symmetric time evolution as a coherent operation} \label{sec:coh}
  Quantum coherence is the ability to form superposition of quantum state. It is usually defined with respect to a fixed basis. Specifically, given a basis $\{\ket{e_i}\}_{i=0}^{d-1}$ for a $d$-level system, a state $\rho$ is \textit{incoherent} if it is diagonal in this basis, i.e., if $\rho = \sum_{j} p_j |e_j \rangle \langle e_j|$, where $p_j$'s form some probability distribution. This motivates for a  natural definition of coherence in terms of the off-diagonal elements of the density matrix
 \begin{equation}\label{eq:Coherence}
 	C(\rho)  = \sum\limits_{i,j (i \ne j)} | \rho_{ij}|,
 \end{equation}
 such that $ 0 \le C(\rho) \le 1$. The extreme points $0$ and $1$ corresponds to the  mixed state and the maximally coherent state, respectively.  This measure is monotonic under incoherent completely positive and trace preserving (CPTP) operations.

			Let $\mathfrak{I}$ denote a set of incoherent states. A density matrix $\delta \in \mathfrak{I}$, if $\delta = \sum_{j} c_j | j \rangle \langle j|$ is diagonal in the basis $\{\ket{j}\}$.  A  general quantum operation is characterized by a set of completely positive and trace preserving operators, known as   Kraus operators \cite{kraus1971general}, denoted here by  $\{\mathcal{K}_i\}_{i=1}^n$. Unitary dynamics can be thought of as a special case with just one Kraus operator. The operation $ \Lambda^{IO}$  is said to be \textit{incoherent} if \cite{shi2017coherence}  
          \begin{equation}\label{eq:IO}
          \Lambda^{IO} [\delta] = \sum_i \mathcal{K}_i \delta \mathcal{K}_i^\dagger \in \mathfrak{I}.
          \end{equation}
          The application of each Kraus operator individually, cannot generate the coherence, i.e., $\mathcal{K}_i \ket{j^\prime} \approx \ket{j^{\prime \prime}}$, for some $\ket{j^\prime}, \ket{j^{\prime \prime}} \in \{\ket{j}\}$ \cite{streltsov2017structure}. Surprisingly, the time evolution generated by non-unitary operator given in Eq. (\ref{eq:U}) turns out to be a coherent operation, generating coherence in a maximally mixed state. This is one of the main observations of this work and is explained in more detail ahead. Further investigation is invited to explore the interplay various quantum resources viz.,  entanglement and coherence and the non-unitary dynamics generated by non-Hermitian Hamiltonians \cite{javid2019qze,javid2019interplay}. A hierarchy of single qubit incoherent operations suitable under different circumstances exists in the literature \cite{shi2017coherence}.

\begin{figure}
    	\centering
    	\includegraphics[width=80mm]{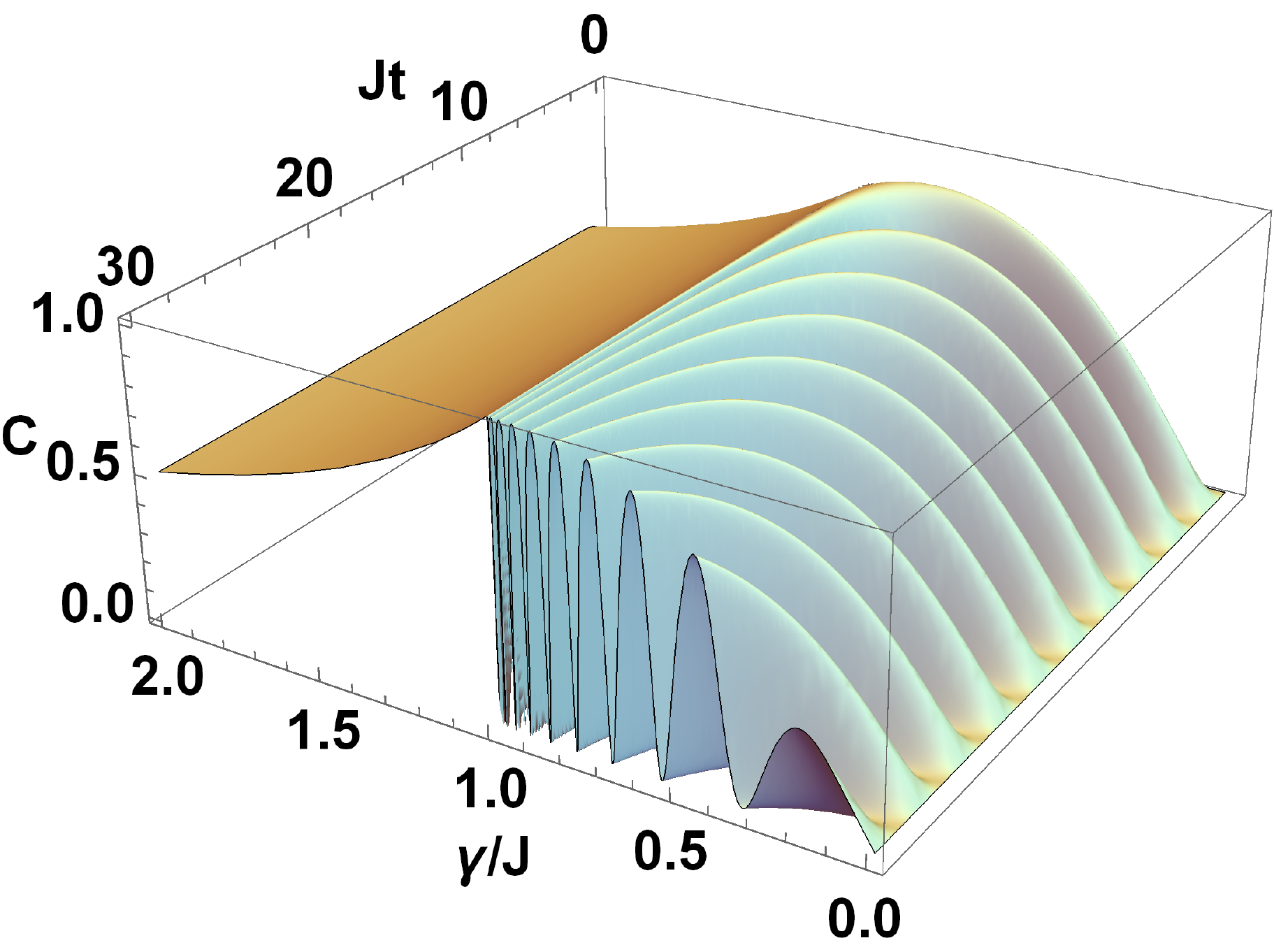}
    	\caption{(Color online) Coherence given in Eq. (\ref{eq:CohExpression}) as a function of dimensionless parameters $ \gamma /J$ and $ J t$.  The coherence attains maximum value around the EP ($\gamma = J$). }
    	\label{fig:Coherence}
    \end{figure}

    The maximally mixed initial state at time $t=0$, i.e., $\rho(0) = \mathcal{I}/2$ ( $\mathcal{I}$ being the identity matrix), evolved for time $t$ according to Eq. (\ref{eq:rhotilde}), leads to the  normalized state, $\tilde{\rho}(t) =\frac{U(t)\rho(0) U^{\dagger}(t)}{\operatorname{Tr}[U(t)\rho(0) U^{\dagger}(t)]}$.  The $l_1$norm of coherence measure for the state $\tilde{\rho}(t)$ turns out to be 
          \begin{equation}\label{eq:CohExpression}
          C(\tilde{\rho} (t)) = 2 \left| \frac{\alpha  \sinh ^2\left( Jt\sqrt{\alpha ^2-1}  \right)}{\alpha ^2 \cosh \left(2 J t \sqrt{\alpha ^2-1}  \right)-1}\right|.
          \end{equation}
             In the limit $\alpha \rightarrow 1$, for large $t$,  $C(\tilde{\rho}(t)) \rightarrow 1$,  the coherence reaches maximum value at the exceptional point. Figure (\ref{fig:Coherence}) depicts the behavior of the coherence parameter $C(\tilde{\rho})$.  It is well known the notion of coherence is intimately related to the mixedness and satisfy the following complementary relation for an arbitrary state $\xi \in \mathbb{C}^d$ \cite{PhysRevA.91.052115} 
          \begin{equation}\label{eq:CohMix}
           \frac{C^2(\xi)}{(d-1)^2} + \mu(\xi) \le 1,
          \end{equation}
          where $\mu(\xi) = \frac{d}{d-1} (1 - \operatorname{Tr}[\xi^2])$ is the mixedness parameter. This relation defines the limits on the degree of coherence imposed by mixedness as a consequence of system environment interactions. This becomes pertinent to the current discussion since the $\mathcal{PT}$ symmetry emerges in our system as consequence of  gain-loss effects.   The corresponding mixedness parameter to the coherence given in Eq. (\ref{eq:CohExpression}) turns out to be		
 
          \begin{equation}
          \mu(\tilde{\rho}) = \frac{\left(\alpha ^2-1\right)^2}{\left(\alpha ^2 \cosh \left(2 Jt \sqrt{\alpha ^2-1}  \right)-1\right)^2}.
          \end{equation}
          In the limiting case with $\alpha \rightarrow 1$, the mixedness parameter $\mu(\tilde{\rho}) = 0$ in accordance with the complementarity relation in Eq. (\ref{eq:CohMix}). Thus, the maximally mixed state subjected to the $\mathcal{PT}$ symmetric dynamics becomes a pure state at the EP, a feature also seen in Fig. (\ref{fig:Coherence}). The unusual enhancement  of coherence about the exceptional points has important consequences; it enables improved quantum violation of some  Leggett-Garg inequalities (LGIs) than that is obtained for the case of unitary evolution. In fact, at EPs, LGIs are violated up to their algebraic maximum. In the next section, we investigate this in detail.
\bigskip

    \section{Consequences of the maximal coherent behavior: Violation of LGIs upto the algebraic maximum}\label{sec:LGIandPT}
		\subsection{Various formulations of LGIs}
		Standard LGIs \cite{leggett85} are derived to test the compatibility between the every-day world view of macrorealism and quantum mechanics. It has two main assumptions:  (i) Macrorealism per se (\textit{MRps}): If a macroscopic system has two or more macroscopically distinguishable ontic states available to it, then the system remains in one of those states at all instant of time (ii) Non-invasive measurability (\textit{NIM}): The definite ontic state of the macrosystem is determined without affecting the state itself or its possible subsequent dynamics. Based on the assumptions of \textit{MRps} and \textit{NIM}, standard LGIs has been derived \cite{leggett85}. Standard LGIs are often considered to be the temporal analog of the Clauser-Horne-Shimony-Halt (CHSH) form \cite{chsh69} of   Bell's  inequalities \cite{bell1964}; however, they are different from the perspective of  measurements. Another interesting difference is the algebraic maximum violation of the respective inequalities. In a local model, the CHSH inequality is upper bounded by $2$ and algebraic maximum is $4$. However, within the standard framework of  quantum mechanics, quantum value of CHSH expression is upper bounded by $2\sqrt{2}$ and cannot reach the algebraic maximum. This inference is valid for dichotomic measurements and is independent of the dimension of the Hilbert space. However, there are post-quantum theories in which the algebraic maximum can be reached. Strong arguments \cite{ic09} are available which show that such a post-quantum correlation (albeit no-signalling) theory does not exist in nature.		
		On the other hand, in \cite{budroni2014} it was shown that the violation of standard LGIs achieve their algebraic maximum for infinitely large system dimension. The conceptual relevance of the results in  \cite{budroni2014} with the standard notion of macrorealism is critically re-examined in \cite{kumari18}. Recently, it has also been shown that the algebraic maximum violation of variants of LGIs can be obtained  even for a qubit system \cite{pan18}. In \cite{kartik19,anant19},  it was brought out that violation of standard LGIs reach up to their algebraic maximum in a qubit undergoing non-Hermitian dynamics. 
		
In light of the above  observations, we investigate the behavior of some of the well studied LGIs under non-Hermitian dynamics governed by Hamiltonian, Eq. (\ref{eq:HPT}).  The standard procedure involves choosing a  dichotomic observable  $\hat{M}$ and computing its two time averages or correlation functions. The assumptions of \textit{MRps} and \textit{NIM} put restrictions on some specific combinations of these two time correlation functions (or the combinations of probabilities).  
The standard LGI involves the computation of the two time correlation function defined as $C_{ij} = \langle \hat{M}(t_i) \hat{M}(t_j) \rangle$, where the average is taken with respect to the state at initial time $t_i$. Considering the measurements of the observable $\hat{M}$ made on macroscopic system at times $t_1$, $t_2$, and $t_3$ ($t_3 > t_2 > t_1$), which in turn implies the measurements of the observables  $M_1$, $M_2$, and $M_3$ respectively, the simplest LGI reads
\begin{equation}\label{eq:K3}
K = C_{12} + C_{23} - C_{13},
\end{equation}
  such that $-3 \le K \le 1$. We will often refer to $K$ as the Leggett-Garg (LG) parameter whose quantum expression  will be denoted by $K_Q$, such that a violation of above inequality  means $K_Q >1$ or $K_Q < -3$ or both. It is well known that in unitary quantum mechanics, the optimal value of LG expression $K_{Q}$ is $1.5$, independent of quantum state.

As discussed above, probabilistic formulations of LGI, for example, the  Wigner form, which is stronger than the standard LGIs,  has been developed. 
They can be derived from the assumptions of joint probability and non-invasive measurability. 
  From the pair-wise statistics of the measurements and by invoking the non-negativity of the probability, Wigner form of LGIs can be obtained as
\begin{eqnarray}
\label{w1}
P(m_{2},m_{3})-P(-m_{1},m_{2})-P(m_{1},m_{3})\leq 0,	
\end{eqnarray}
\begin{eqnarray}
\label{w2}
P(m_{1},m_{3})-P(m_{1},-m_{2})-P(m_{2},m_{3})\leq 0,
\end{eqnarray}
\begin{eqnarray}
\label{w3}
P(m_{1},m_{2})-P(m_{2},-m_{3})-P(m_{1},m_{3})\leq 0.
\end{eqnarray}
One can obtain $24$ Wigner form of LGIs from inequalities (\ref{w1}-\ref{w3}).
 
It has  recently been shown that Wigner form of LGIs are not only inequivalent but also stronger than the standard LGIs \cite{pan17,swati17,pan19}. This inequivalence can be shown using the moment expansion of pair-wise probability in quantum theory. We refer interested reader to Ref.\cite{pan19,javid20} for detailed discussion. Now, we probe standard and Wigner form of LGIs in $\mathcal{PT}$ symmetric dynamics.

\subsection{Quantum violation of LGIs}
For our purpose, we choose the qubit observable $\hat{M} = \sigma_y$, the Pauli-$y$ matrix. The same observable will be measured at three different times $t_1$, $t_2$ and $t_3$. Initializing our system, at $t=0$, the maximally mixed state  $\rho(0)=\mathcal{I}/2$, the state evolves according to Eq. (\ref{eq:rhotilde}) giving  $\tilde{\rho}(t) =\frac{U(t)\rho(0) U^{\dagger}(t)}{\operatorname{Tr}[U(t)\rho(0) U^{\dagger}(t)]}$. 
\begin{figure*}
\hspace{1cm}\includegraphics[width=65mm]{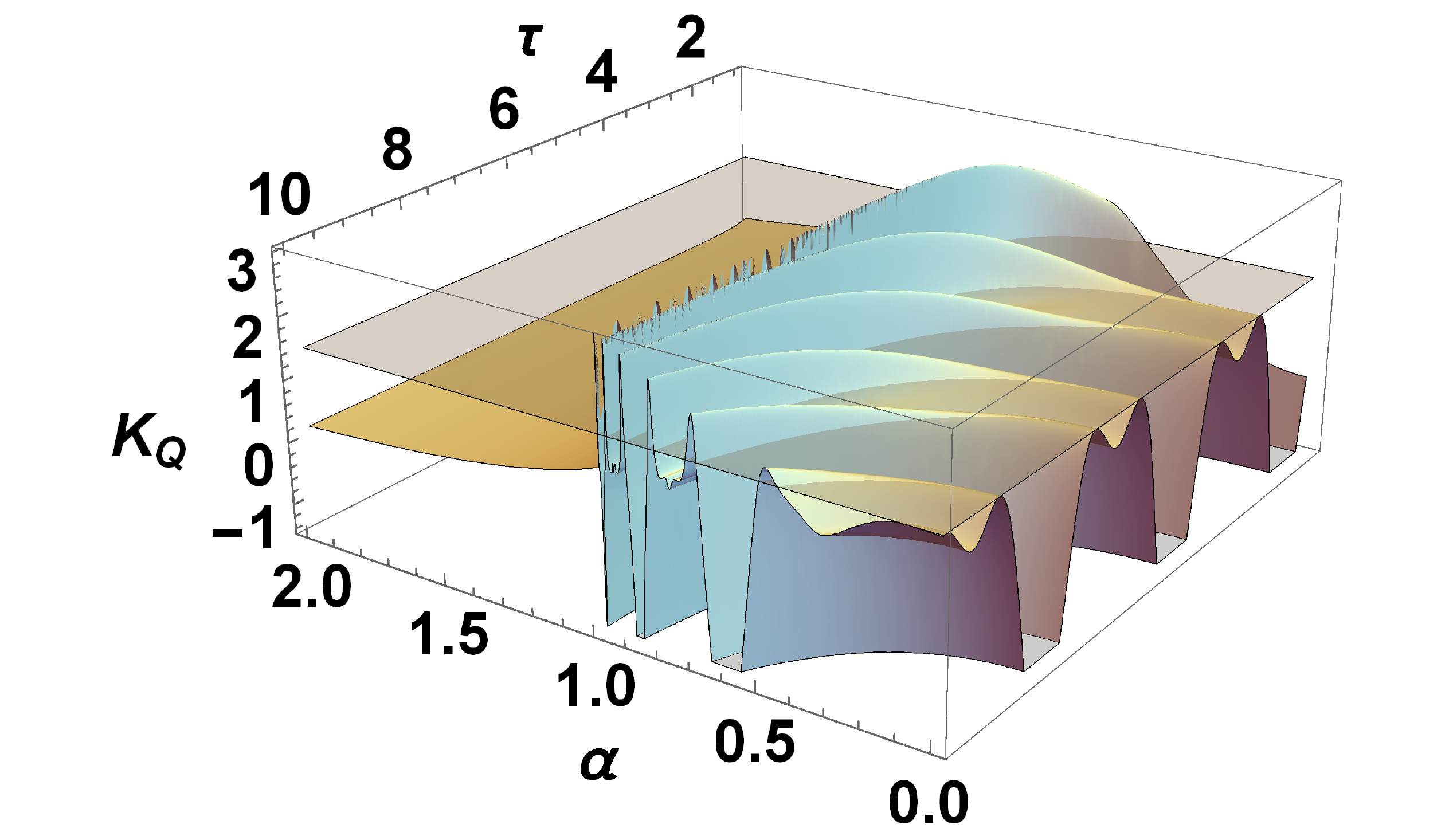}\hspace{1cm}
	\includegraphics[width=65mm]{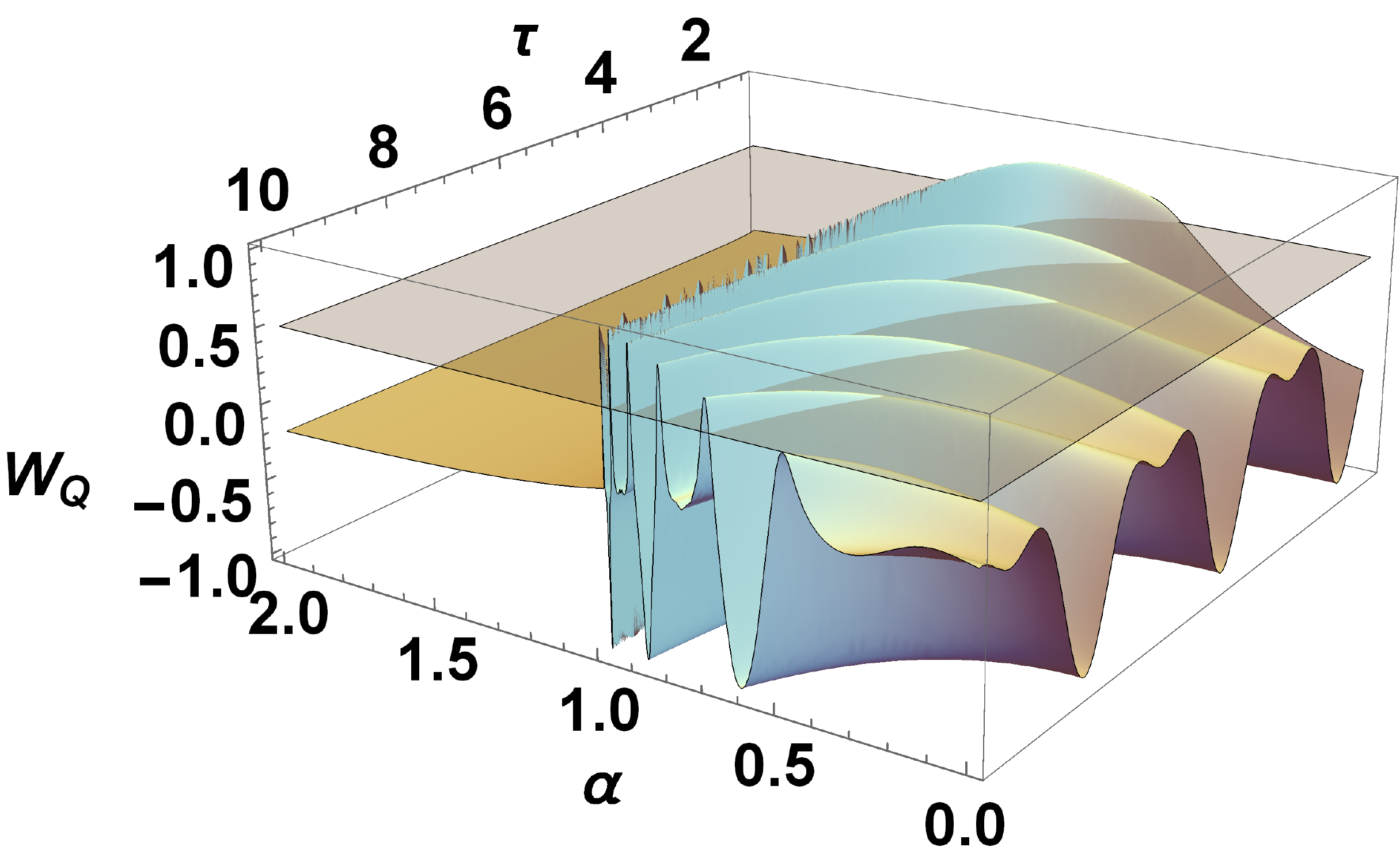}\\
	
	\hspace{1cm} \includegraphics[width=60mm]{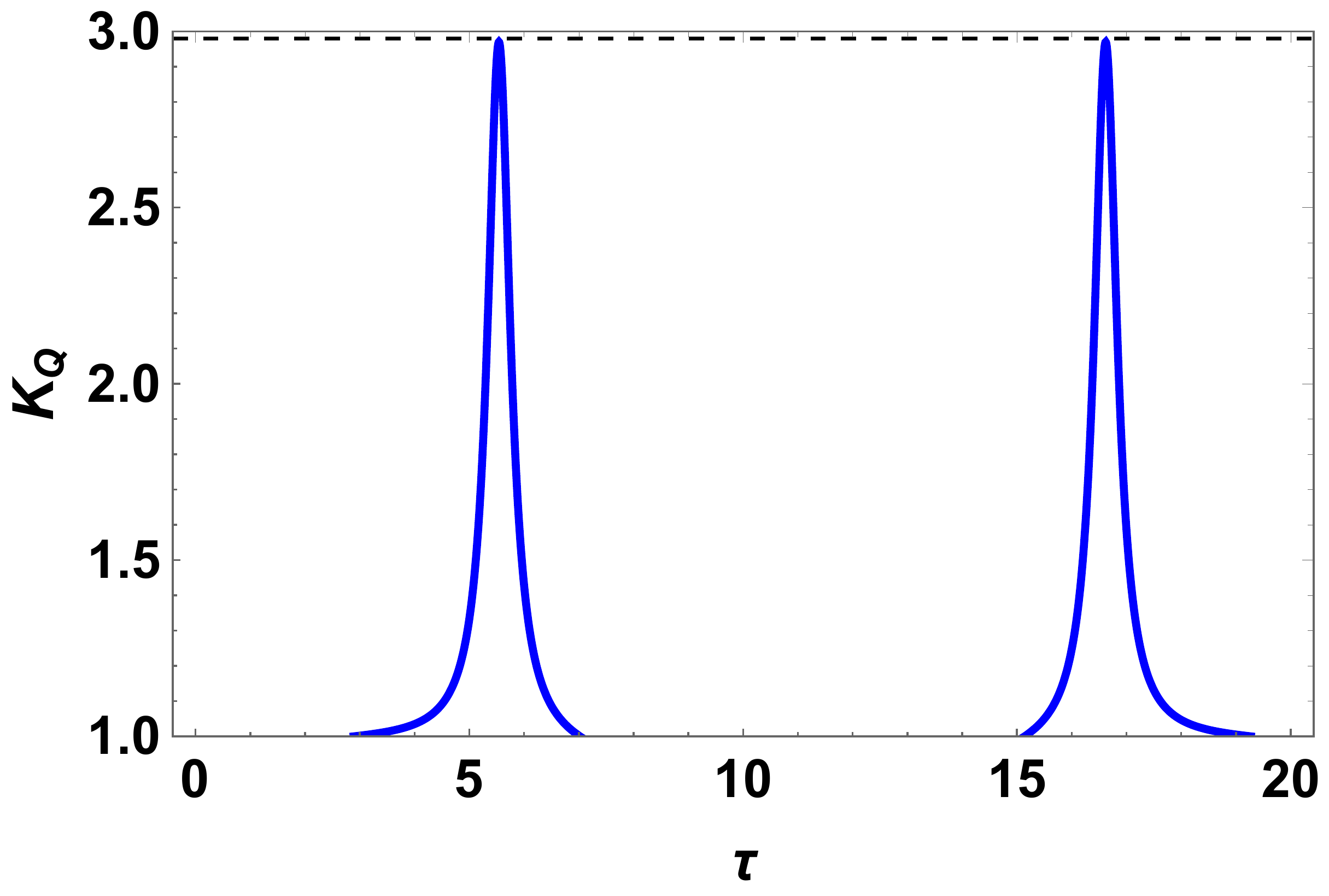}\hspace{2cm}
	\includegraphics[width=60mm]{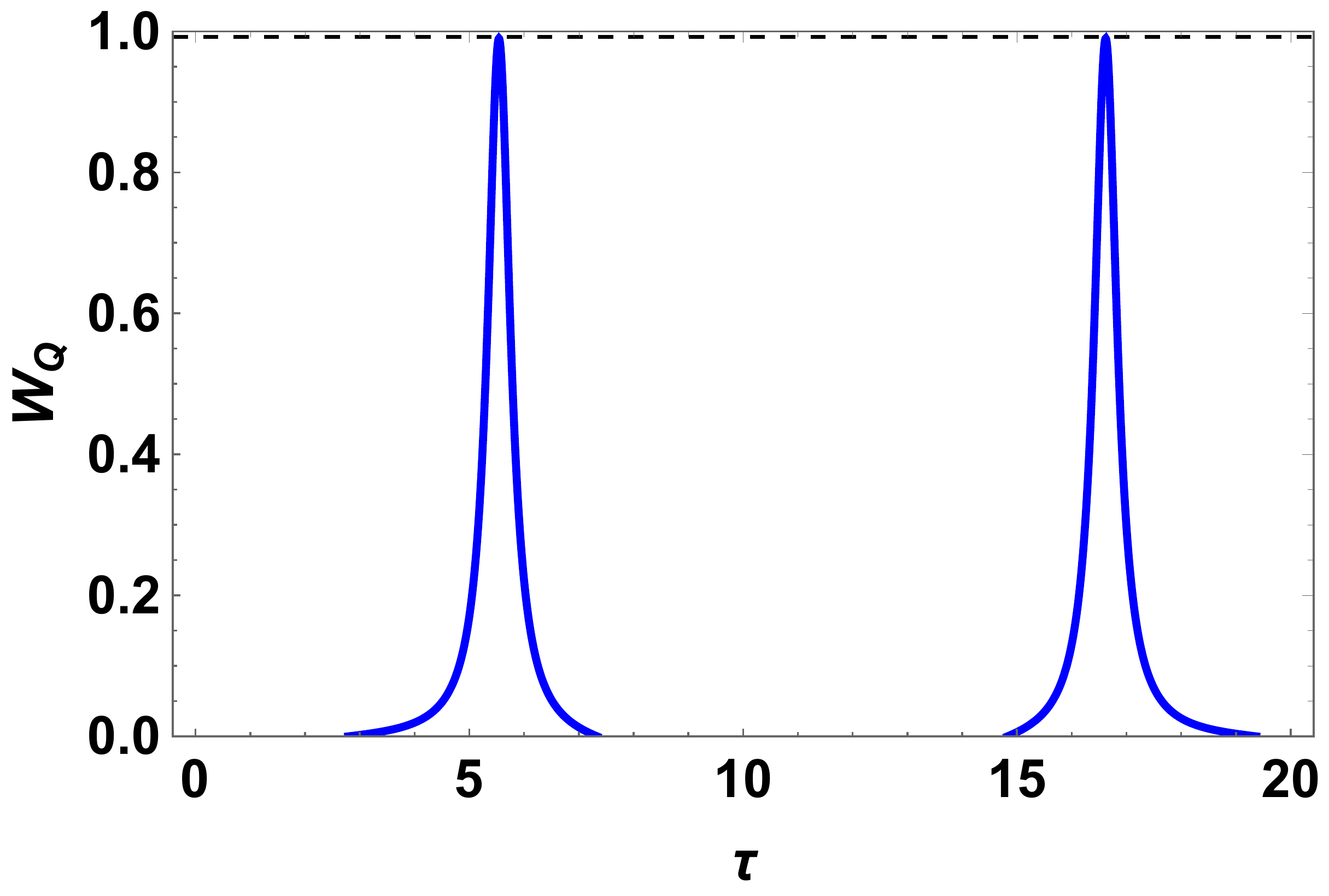}	
	\caption{(Color online) Standard LGIs as defined in Eq.(\ref{eq:K3}) and Wigner (inequality(\ref{w1})) form of LGIs are plotted with respect to the dimensionless parameter $\uptau = J (t_2 - t_1) = J (t_3 - t_2)$ and $\alpha = \gamma/J$. The plots in the bottom panel bring out, in a more clear manner, the behavior at the EP, i.e., $\alpha = 1$.  At this point  LGIs achieve their algebraic maximum.}
	\label{fig:K3W3}
\end{figure*}

For the model considered here, the quantum expression of the standard LG expression for Eq. (\ref{eq:K3}) turns out to be
	\begin{equation}\label{eq:K3expr}
             K_Q   = \langle \sigma_y(t_1) \sigma_y(t_2) \rangle + \langle \sigma_y(t_2) \sigma_y(t_3) \rangle - \langle \sigma_y(t_1) \sigma_y(t_3) \rangle.
\end{equation} 
The two time correlations $\langle \sigma_y(t_i) \sigma_y(t_j)\rangle $ can be expressed in terms of joint probabilities $ p(^{a}t_i, ^{b}t_j)$ of obtaining outcome $a$ at $b$ at times $t_i$ and $t_j$, respectively, with $a,b \in \{\pm 1\}$ 
\begin{equation}
\langle \sigma_y(t_i) \sigma_y(t_j) \rangle = \sum_{a, b = \pm 1} a b ~ p(^{a}t_i, ^{b}t_j).
\end{equation} 
The joint probabilities can be computed as 
\begin{equation}\label{eq:JointProb}
p(^{a}t_i, ^{b}t_j) = \operatorname{Tr} \{ \Pi^{b} U(t_j - t_i) \Pi^{a} \rho(t_i) \Pi^{a} U^\dagger(t_i-t_j) \}.
\end{equation}

Here, $\rho(t_i) = U(t_i) \rho(0) U^\dagger (t_i) / \operatorname{Tr}[U(t_i) \rho(0) U^\dagger (t_i)]$ is the normalized state at time $t_i$.
By using Eqs. (\ref{eq:U}) and (\ref{eq:JointProb}), the quantum LG expression $K_Q$ from Eq. (\ref{eq:K3expr}) can be calculated as
\begin{equation}
K_Q  =   \sum\limits_{n = 0}^{9}  \frac{1}{N}c_n \cosh (2n \Theta).
\end{equation}
with
\begin{equation}
\Theta = \uptau \sqrt{\alpha^2 - 1}, \quad \uptau = J(t_2 - t_1) = J(t_3 - t_2),
\end{equation}
  and
\begin{align}
c_{0} &= 16 \alpha^2  + 26 \alpha^4 - 44 \alpha^6 + 12 \alpha^8,\nonumber \\
c_{1}&=  64 - 104 \alpha^2 + 20 \alpha^4 + 28 \alpha^6 - 14 \alpha^8, \nonumber \\
c_{2} &= 4(-8-8\alpha^2 +23 \alpha^4 - 15 \alpha^6 + 6 \alpha^8), \nonumber \\
c_{3} &= 2 (-16 + 24 \alpha^2 + \alpha^4 - 5\alpha^6), \nonumber \\
c_{4} &= 8(6 - 3 \alpha^2 - 6 \alpha^4 + 2 \alpha^6), \nonumber \\
c_{5} &= -24 + 16 \alpha^2 + 14 \alpha^4, \nonumber \\
c_{6} &= 4 \alpha^2 - 4 \alpha^4 + 8 \alpha^6,\nonumber \\
c_{7} &= 12 \alpha^2 - 13 \alpha^4 - 2 \alpha^6,\nonumber \\
c_{8} &=-2 \alpha^2 - 4 \alpha^4 + 4 \alpha^6, ~~ c_{9} = \alpha^4.
\end{align}
Also,
\begin{align}
N &= (-1 + \alpha \cosh2 \Theta )(-1 + \alpha \cosh4 \Theta) \nonumber \\ &\times (-1 + \alpha^2 \cosh 2 \Theta) (-1 + \alpha^2 \cosh 4\Theta) \nonumber \\& \times ( 1 + \alpha \cosh 2\Theta)(1 + \alpha \cosh4\Theta),
\end{align}
 which arises from time normalization of state in Eq. (\ref{eq:JointProb}) in the calculation of $K_Q$.  As already mentioned that the optimal value of $K_{Q}$ in unitary quantum theory is upper bounded by $1.5$.  We show that in $\mathcal{PT}$ symmetric  evolution the value of $K_{Q}$ can be beyond this limit. As an example, for the values of $J=0.6$, and $\gamma=0.5$ , i.e., $\alpha=0.9$ (hence, in the $\mathcal{PT}$ symmetric region), the maximum value of $K_{Q}$ is found to be $2.54$. However, when $J \rightarrow \gamma$ ($\alpha \rightarrow 1$),   the quantum value of standard LGI approaches to its algebraic maximum $3$, as depicted in  Figure \ref{fig:K3W3}.

\begin{figure}
	\centering
	\includegraphics[width=80mm]{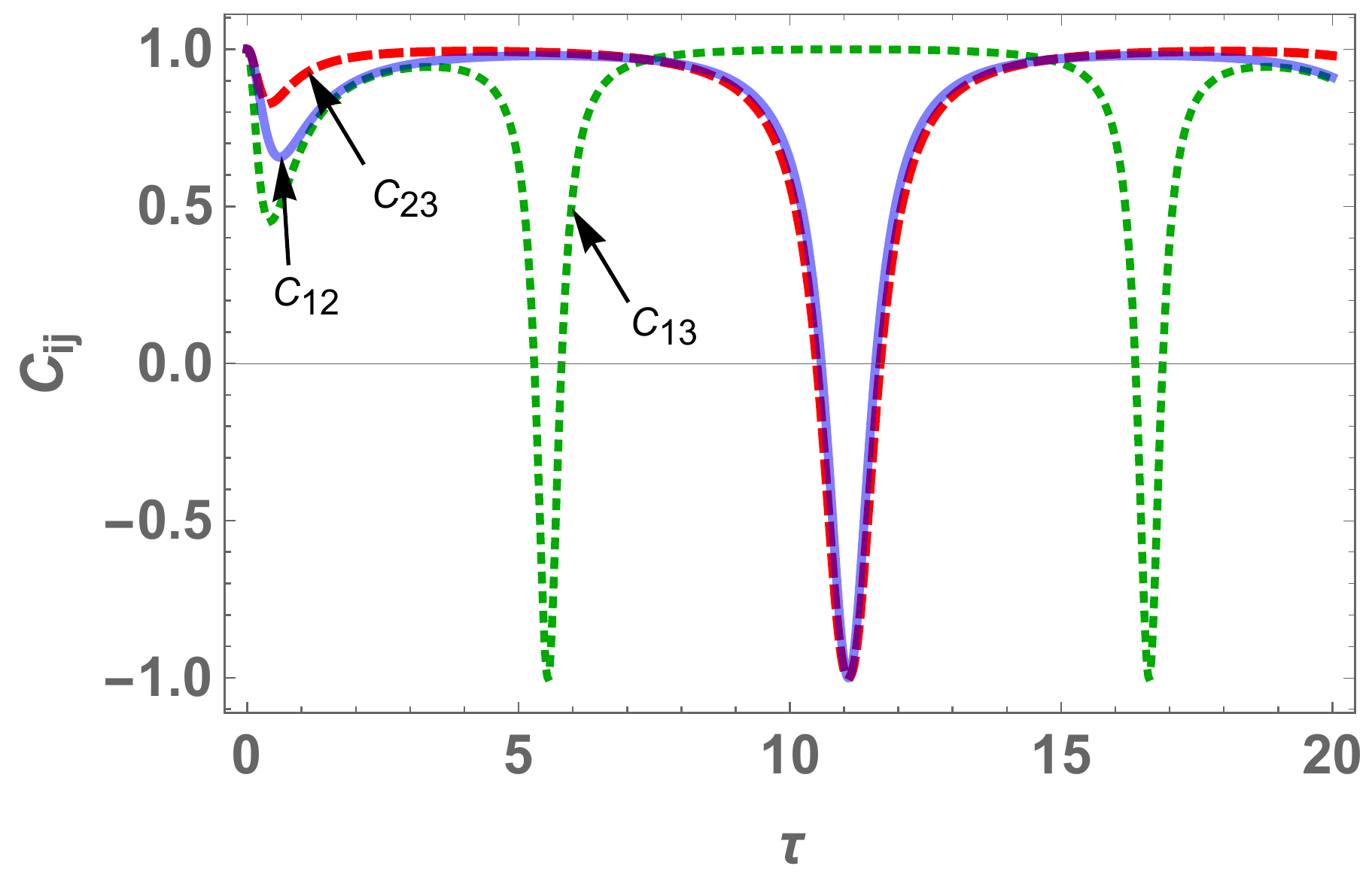}
	\caption{~ (Color online) Depicting the two time correlation functions $C_{ij} = \langle \sigma_y(t_i) \sigma_y(t_j) \rangle$ with respect to (dimensionless) time separation $\uptau = J (t_j - t_i)$ around the excetional point.}
	\label{fig:Cij}
\end{figure}

 The algebraic maximum can be obtained if and only if first two correlations in Eq. (\ref{eq:K3expr}) are equal to $+1$ and the third correlation is $-1$. Now, a mixed state $\rho(0)=\mathcal{I}/2$ subjected to the dynamics generated by the time evolution operator in Eq. (\ref{eq:U}), becomes a pure state at a later  time $t_1$, which is an eigenstate of $\sigma_y$ with eigenvalues $-1$. A measurement is now made at time $t_1$ and the post-measurement state is evolved to time $t_2$. At the EP, this state coincides with the state at time $t_1$ leading to $\langle \sigma_y(t_1)\sigma_y(t_2)\rangle=1$. Same explanation holds for $\langle \sigma_y(t_2)\sigma_y(t_3)\rangle=1$. However, the time evolution of the post-measurement state at $t_1$ to some later time   $t_3$, such that $t_3 - t_1 = 2 (t_2-t_1)$, leads to the state of the system which is  an eigenstate of $\sigma_y$ with eigenvalue $+1$. Consequently, the correlation $\langle \sigma_y(t_1)\sigma_y(t_3)\rangle=-1$, and the sum of all the three correlations reaches the algebraic  maximum of $3$.  This is corroborated by  Fig. (\ref{fig:Cij}) which depicts the variation of the two time correlation functions with respect to the time separation $\uptau$.

 Next, in order to study the Wigner form of LGI, we make use of inequality  (\ref{w1}), which represents a set of eight inequalities for different combinations of values $m_1, m_2, m_3 \in \{\pm 1 \}.$ We choose a particular inequality for which  $m_1=+1$, $m_2=-1$ and $m_3=-1$, that is 
\begin{align}
W &= P(m_{2}=-1,m_{3}=-1)-P(-m_{1}=-1,m_{2}=-1)  \nonumber \\&-P(m_{1}=+1,m_{3}=-1)\leq 0.
\end{align}
Using Eq. (\ref{eq:JointProb}) to calculation the joint probabilities,  the quantum expression of Wigner form of LGI, which we denote by  $W_Q$, is given by 
\begin{eqnarray}
			\label{wq}
		W_Q &=\frac{1}{2}\big[\frac{(\alpha +1)^2 \left(\alpha  \cosh \left(4 \uptau \sqrt{\alpha ^2-1} \right)-1\right) \cosh ^2\left(\uptau\sqrt{\alpha ^2-1}  \right)}{\left(\alpha  \cosh \left(2  \uptau \sqrt{\alpha ^2-1}\right)+1\right) \left(\alpha ^2 \cosh \left(4 \uptau\sqrt{\alpha ^2-1}  \right)-1\right)}\nonumber\\
		&-\frac{(\alpha +1)^2 \cosh ^2\left(2 \uptau\sqrt{\alpha ^2-1}  \right) \left(\alpha  \cosh \left(2 \uptau\sqrt{\alpha ^2-1}  \right)-1\right)}{\left(\alpha ^2 \cosh \left(2 \uptau\sqrt{\alpha ^2-1}  \right)-1\right) \left(\alpha  \cosh \left(4 \uptau\sqrt{\alpha ^2-1}  \right)+1\right)}\nonumber\\
		&-\frac{\left(\alpha ^2-1\right) \sinh ^2\left(\uptau\sqrt{\alpha ^2-1}  \right) \left(\alpha  \cosh \left(2 \uptau\sqrt{\alpha ^2-1}  \right)+1\right)}{\left(\alpha  \cosh \left(2 \uptau\sqrt{\alpha ^2-1}  \right)-1\right) \left(\alpha ^2 \cosh \left(2 \uptau\sqrt{\alpha ^2-1}  \right)-1\right)}\big].
		\end{eqnarray}
		For the same values of the parameters in $\mathcal{PT}$ symmetry region considered in the standard LG case, i.e., $J=0.6$ and $\gamma=0.5$, the maximum quantum violation of Wigner form of LGIs is obtained as $0.84$. At the EP $(\alpha=1)$, the optimal value of $W_Q$, given by Eq. (\ref{wq}), achieves the algebraic maximum $1$.  Thus, the violation of both the standard and the Wigner form of LGIs  reaches up to the algebraic maximum,  as  depicted in Fig. (\ref{fig:K3W3}). Taken in conjunction with the behavior of coherence at the EP, it emerges that  in $\mathcal{PT}$ symmetric dynamics, the maximal enhancement of coherence near the exceptional point leads to the violation of various formulations of LGIs up to the algebraic maximum.\bigskip

\section{Summary and Conclusion}\label{sec:SC}
The $\mathcal{PT}$ symmetric systems exhibit intriguing behavior around the exceptional points (EPs)$-$the points of coalescence of both eigenvalues as well as  eigenvectors.  The  EPs differ from the diabolic points in Hermitian systems where only eigenvalues show coalescence.  We started with a non-Hermitian  $\mathcal{PT}$ symmetric system and showed that it can be viewed as a Hermitian system in a higher dimensional Hilbert space using Nimark dilation theorem.  Interestingly, it is found that a maximally mixed state acquires coherence when subjected to such  non-Hermitian dynamics.  The $l_1$ square norm based measure of coherence exhibits very distinct behavior in  $\mathcal{PT}$ symmetric and  $\mathcal{PT}$ symmetry broken phases with recurrent behavior in the former case, Fig.(\ref{fig:Coherence}). Near  the EPs, the coherence shows unconventional enhancement and reaches its maximum value, obeying  the complementary relation of  coherence and  mixedness parameters  at this point.

The enhancement in the coherence about EPs can have  interesting consequences. We show its impact on the degree of violation of Leggett-Garg inequalities (LGIs). The quantum violation of various formulations of LGIs not only exceed their quantum bound for qubit but also reach their algebraic maximum at EPs.  In the absence of gain/loss $\gamma=0$, the LG parameter given in Eq. (\ref{eq:K3expr}) reduces to $ K_Q =  2 \cos 2  \uptau - \cos 4 \uptau$, which attains a maximum of $3/2$ (the maximum quantum bound) for a qubit for $ \uptau = \pi /6$.

 	  Recently, efforts have been made to use LGIs to identify the order-disorder quantum phase transitions \cite{gomez2016quantum}, characterization of quantum transport \cite{lambert2010distinguishing} and to distinguish the topological phase transitions \cite{gomez2018universal}.  The current work adds to the list  of nontrivial phenomena  about the EPs, and can have potential applications in carrying out tasks where quantum coherence plays a fundamental role.   The present work may be extended to the higher dimensional discrete spaces. It would also be interesting to look for the interplay of $\mathcal{PT}$ symmetry and LGI violations in the context of continuous infinite dimensional systems \cite{javid2019qze}. Futher, alternative methods for dealing with non-Hermitian systems  like Krein space formalism  \cite{mostafazadeh2006krein} can be used instead of the effective Hamiltonian approach.\bigskip

 	  \acknowledgments
 JN's  work was supported by the project  ``Quantum Optical Technologies" carried out within the International Research Agendas programme of the Foundation for Polish Science co-financed by the European Union under the European Regional Development Fund. SK is supported by the Ministry of Science \& Technology (Grant no. 109-2811-M-006 -513). 	AKP acknowledges the support from the project DST/ICPS/QuEST/2018/Q-42.
 	
%
\end{document}